\documentclass{moriond}

\def\Journal#1#2#3#4{{#1} {\bf #2}, #3 (#4)}

\def\PRD{{\em Phys. Rev.} D}

\def\be{\begin{equation}}
\def\ee{\end{equation}}
\def\bea{\begin{eqnarray}}
\def\eea{\end{eqnarray}}

\newcommand{\Photo}{}

\begin{document}
\vspace*{4cm}
\title{Progress on testing Lorentz symmetry with MICROSCOPE}
\author{H. Pihan-Le Bars$^1$, C. Guerlin$^{2, 1}$, P. Wolf$^1$\\
\ \\}
\address{$^1$ SYRTE, Observatoire de Paris, PSL Research University, CNRS, Sorbonne Universit\'es, UPMC
Univ. Paris 06, LNE, 61 avenue de l'Observatoire, 75014 Paris, France}
\address{$^2$ Laboratoire Kastler Brossel, ENS-PSL Research University, CNRS, UPMC-Sorbonne Universit\'es,
Coll\`ege de France}
\maketitle\abstracts{
The Weak Equivalence Principle (WEP) and the local Lorentz invariance (LLI) are two major assumptions of General Relativity (GR). 
The MICROSCOPE mission, currently operating, will perform a test of the WEP with a precision of $10^{-15}$. The data will also be analysed at SYRTE for the purposes of a LLI test realised in collaboration with J. Tasson (Carleton College, Minnesota) and Q. Bailey (Embry-Riddle Aeronautical University, Arizona).
This study will be performed in a general framework, called the Standard Model Extension (SME), describing Lorentz violations that could appear at Planck scale ($10^{19}$ GeV). The SME allows us to derive a Lorentz violating observable designed for the MICROSCOPE experiment and to search for possible deviations from LLI in the differential acceleration of the test masses~\cite{tasson2011}.}

\section{Lorentz violation in SME framework}
\subsection{Generalities}
The LLI can describe the invariance under coordinate changes, called observer invariance, as well as the invariance under orientation and boost changes in a given observer frame - the particle invariance. In case of a space-time anisotropy and in the context of the SME framework, the observer invariance is preserved, but the particle invariance can possibly be violated.\par 
The SME is a general framework built to address the issue of low-energy effects of new physics that could appear at Planck scale in both GR and Standard Model (SM), and particularly spontaneous Lorentz symmetry breaking~\cite{colladay1998}. It is an effective theory built from the combination of Standard Model (SM) fields and considering a curved spacetime. 
The resulting additional Lorentz violating terms appearing in the Lagrangian are parametrized by SME background coefficients (scalars, vectors or tensors) quantifying the amplitude of the deviation to the symmetry in a large variety of sectors (e.g. photon, matter, gravity). Most of these coefficients are composition-dependent, meaning that they lead to different deviations to known physics depending on the considered body or particle.
\begin{equation}
\mathcal{L}_{photon} = \frac{1}{4} F_{\mu\nu} F^{\mu\nu} -  \frac{1}{4} (k_F)_{\kappa \lambda \mu\nu} F^{\kappa\lambda} F^{\mu\nu}
\label{lag}
\end{equation}
The Eq.~\ref{lag} illustrates an SME modification to the pure-photon sector Lagrangian \cite{colladay1998}. The first term is the usual Lagrangian for Maxwell's equation in absence of sources, and the second term is an SME violating term constructed from the standard Maxwell tensor and parametrized by the SME coefficient $(k_F)$ which is a rank-4 tensor.
All these coefficients are free parameters of this effective theory and need to be constrained experimentally. 

\subsection{SME observables}
\label{12}
The SME intends to be independent of any given theory leading to such Lorentz violation, which enables the derivation of general observables for a large range of experiments that cannot be easily done in such theories. However the lack of assumptions on the underlying mechanism for the symmetry breaking makes the SME unable to predict the amplitude of the Lorentz violation signals arising from the coupling between dynamics and SME background coefficients.\par
\begin{figure}[h!]
\centering
 \includegraphics[width =0.6 \linewidth]{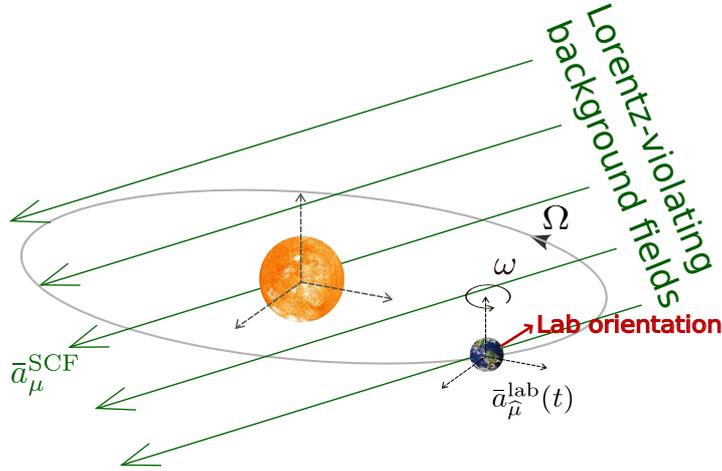}
   \caption{Schematic view of the trajectory of a Earth based lab frame with respect to the SCF. The vector field $\bar{a}_\mu$ is an arbitrary SME background coefficient. The Earth's orbital and sidereal frequencies are respectively denoted $\Omega$ and $\omega$.}
   \label{fig1}
\end{figure}

Moreover, SME coefficients are coordinate-dependent, but are supposed to be constant in a cosmological frame as well as in any frame having a rectilinear motion with respect to it, \textit{e.g.} the Sun Centered Frame (SCF) which is the conventional frame used to report the results of SME analyses. For a ground or a space based experiment, the transformation between the moving lab frame and the SCF leads to a time-dependent observable whose amplitude is a function of constant SCF SME coefficients, in contrast to the lab frame observable which is function of time-varying lab frame coefficients. This SCF observable can be composed of several harmonics, each having its proper amplitude and dependence on SME coefficients, and whose frequencies are related to the orbital motion of the lab frame with respect to the SCF (see Fig.~\ref{fig1}).


\section{Application to MICROSCOPE experiment}

\begin{figure}
\begin{minipage}{0.3\linewidth}
\centerline{\includegraphics[height=0.19\textheight]{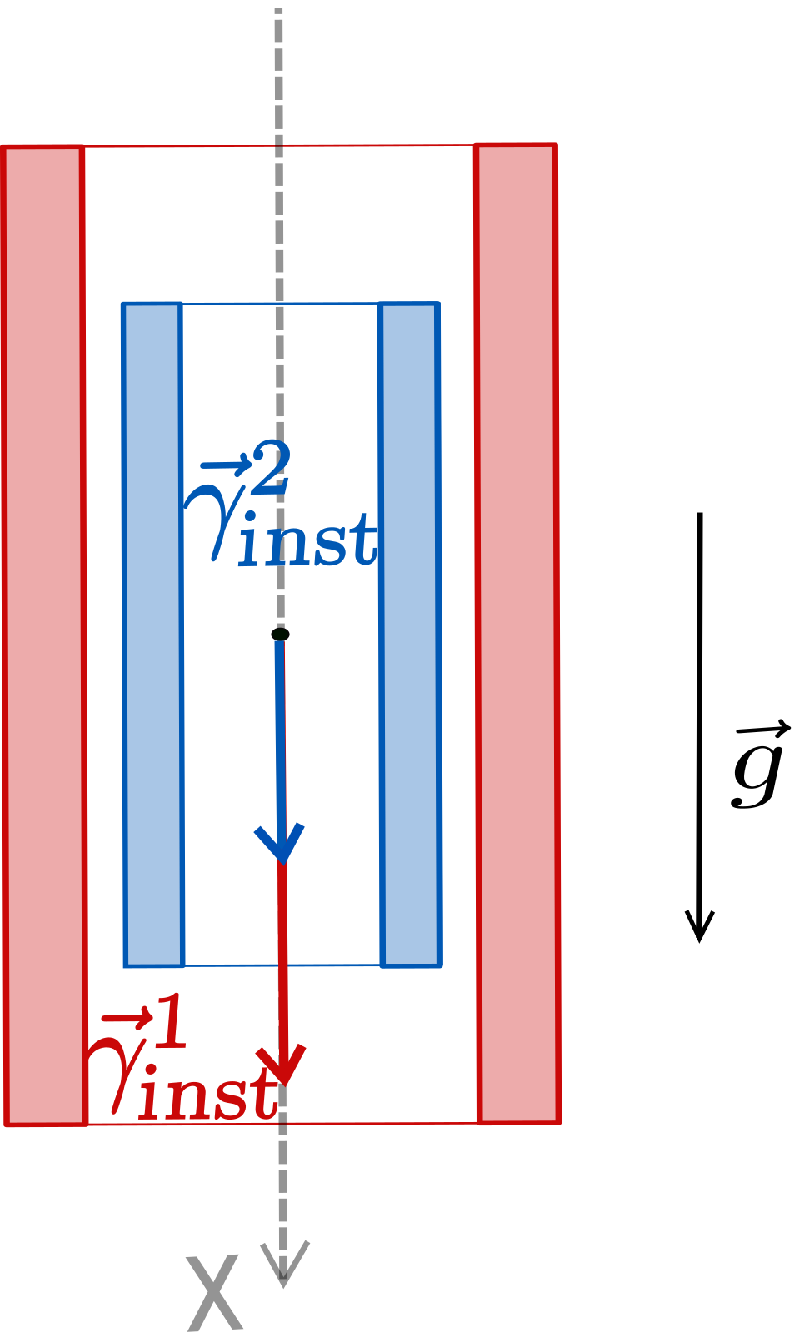}}
\end{minipage}
\begin{minipage}{0.3\linewidth}
\centerline{\includegraphics[height=0.1\textheight]{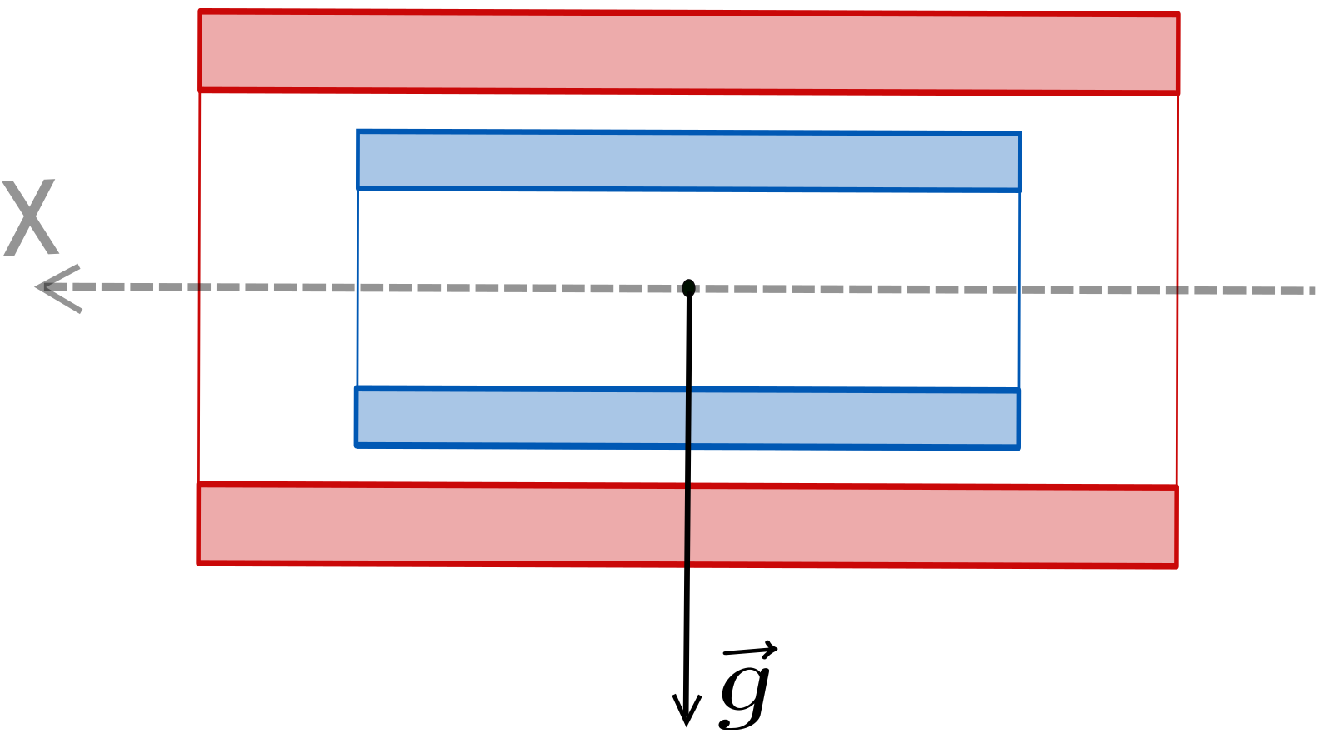}}
\end{minipage}
\hspace{3mm}
\begin{minipage}{0.3\linewidth}
\centerline{\includegraphics[height=0.22\textheight]{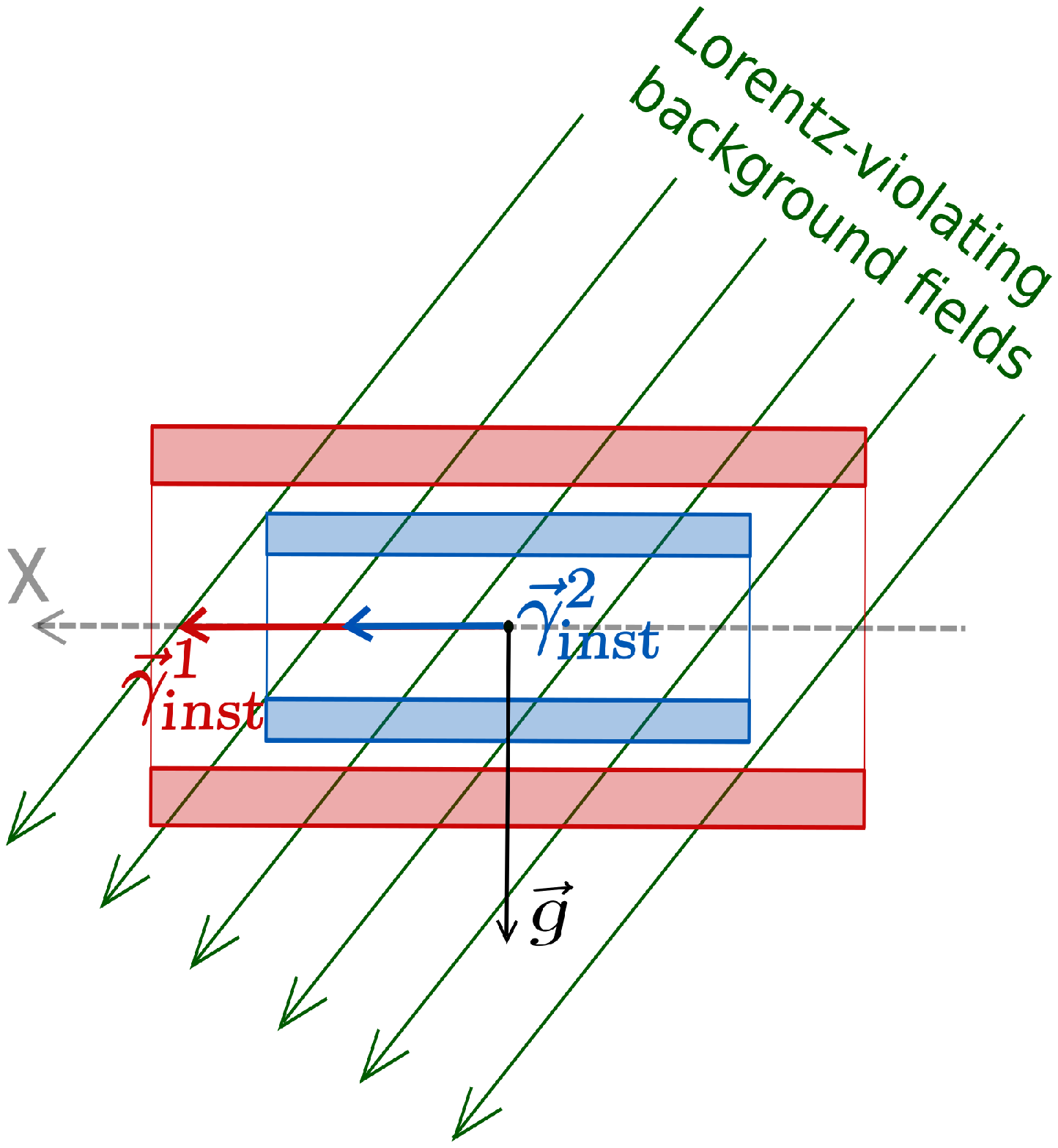}}
\end{minipage}\\
\  \\
\begin{flushleft}
\hspace{0.8cm} (a) Signal $\neq 0$\hspace{3cm} (b) No signal \hspace{2.5cm} (c) Signal $\neq 0$
\end{flushleft}
\caption[]{Schematic view of the test masses: the blue one is in platinum and red one is the titanium mass. The acceleration of the titanium and platinum test masses in the instrument frame are respectively denoted $\vec{\gamma}^1_{\mathrm{inst}}$ and $\vec{\gamma}^2_{\mathrm{inst}}$. Cases (a) and (b) illustrate two scenarii of the WEP test: in (a) the sensitive axis $X$ is parallel to the gravitational acceleration $\vec{g}$ and in (b) it is perpendicular to $\vec{g}$. Case (c) shows the SME equivalent to (b), where a violation signal can be expected.}
\label{fig2}
\end{figure}

Several configurations that can be encountered in the WEP and SME LLI tests are described on Fig.~\ref{fig2}. Cases (a) and (b) illustrate two typical scenarii of the WEP test: one where a violation signal could be measured (a) and one where no signal is expected (b). Configuration (a) could as well involve a violation signal in the LLI test, despite another signature, but a SME violation could also be expected when the gravitational acceleration $\vec{g}$ is perpendicular to the sensitive axis, as shown with case (c). This is due to the coupling between SME background coefficients and dynamics, such signal is specific to the SME LLI test. The performance of the SME test can easily be compared with the WEP test thanks to an analytical matching between the E\"otv\"os parameter $\eta$ and a component of a Geocentric Celestial Reference Frame (GCRF) SME coefficient, $(\bar{a}_{\mathrm{eff}})_t$. \par 
The nature of the Lorentz violation occurring in the SME leads to harmonics that are not expected in the WEP test, which can only involve a violation signal at frequencies $\omega_{o}\pm \omega_{s}$ with $\omega_{o}$ the satellite orbital frequency and $\omega_{s}$ the satellite spinning frequency. In the SME case, a violation could be expected at frequencies $\lbrace  \omega_{s}, \omega_{o},  \omega_{s}\pm \omega_{o},  \omega_{s}\pm 2 \omega_{o},   2\omega_{s}\pm2 \omega_{o},  \omega_{s} \pm \omega_{o}\pm \Omega \rbrace $~\cite{tasson2011}, with $\Omega$ the Earth's orbital frequency.\par

The main systematics that could overlap with SME violation signals will be due to offcenterings, which are mainly expected to contribute at frequency $2\omega_o$. However, even if some SME harmonics are strongly correlated with offcenterings, SME coefficients could still be decorrelated from each other and from offcenterings as the SME test involves many frequency components with amplitudes ruled by a particular combination of SME coefficients for each of them. During our analysis, we will investigate this correlation between SME coefficients and offcenterings.\par 
As we are measuring a differential acceleration, only composition-dependent coefficients can be constrained with the MICROSCOPE observable: the rank-2 tensor $\bar{c}_{\mu\nu}$ and the vector $(\bar{a}_{\mathrm{eff}})_\mu$, with coordinates denoted by $x_\mu=\left( t,x,y,z\right)$. They both lead to different trajectories for the platinum and the titanium mass. Based on the attempted performance of the mission, it has been estimated that we could improve by up to 3 orders of magnitude the limits on some of these coefficients~\cite{tasson2011}, but it will depend on the noise and systematics encountered during the flight.

\section{Status}
\subsection{SME observable for MICROSCOPE}
We chose a semi-analytical approach to obtain the MICROSCOPE instrument observable, \textit{i.e.} the instrument frame acceleration depending on SCF SME coefficients: first through an analytical derivation of the differential acceleration in the instrument frame in terms of GCRF coefficients, and then a numerical instantaneous transformation from GCRF to SCF coefficients.\par 
The first step is based on the Lagrangian introduced in~\cite{tasson2011} and expressed in the GCRF. In order to determine the instrument differential acceleration, one needs to change the GCRF acceleration - obtained from Euler-Lagrange equations - into the instrument acceleration through a coordinate transformation. After this process, the instrument acceleration is still a function of GCRF coefficients, which enables comparisons with WEP test, keeping only the $(\bar{a}_{\mathrm{eff}})_t$ coefficient in the SME observable.\\
The instantaneous transformation from GCRF to SCF coefficients will be performed numerically using INPOP ephemerides provided by IMCCE. This last operation is needed to express the observable in terms of SCF coefficients, which is conventional in SME tests and introduces new violation frequencies in the model, as explained in Sec.~\ref{12}.\par 
The first expression for the LLI violating observables in Microscope was given in terms of Keplerian elements~\cite{tasson2011}. Here we designed our model according to simulated data provided by ONERA/OCA in order to include as far as possible the in-flight parameters. This in combination with the semi-analytical approach makes our model flexible and convenient for the comparison with the WEP test.
\subsection{Data analysis}
For now, we worked on a first set of simulated data provided by ONERA/OCA. This enabled us to familiarise ourselves with the data and to perform some least-squares fitting on toy-models. Among other things, we identified that the contribution of the Earth's Newtonian gravitational potential, initially taken in the approximation of a uniform spherical source~\cite{tasson2011}, needs to take into account higher order harmonics in order to allow direct comparison with results obtained in the standard WEP analysis. We also gradually including the contributions from angular velocity matrix, gravity gradient matrix, and so on.\par 
The first toy-models adjustments were consistent with the parameters used in the simulations, however it has been done using unrealistic data, without noise. We are currently working on the same kind of test with new sets of simulated data including a realistic noise. Obviously, this requires the implementation of statistical methods to deal with this realistic coloured  noise, which prevents the use of a simple least-squares fitting. As a first step, we are planning to use Monte-Carlo simulations based on a noise model provided by ONERA/OCA to perform the statistical analysis.

\section{Conclusion}
The preparation of the SME analysis of MICROCOPE data is currently ongoing at SYRTE. We are working on the derivation of an observable designed according to the in-flight parameters. The implementation of statistical methods for the data analysis is also in progress, thanks to simulated data provided by ONERA/OCA.\par
In the next few months, we will increase the complexity of the model and of the simulated data in order to check the consistency of both the observable and the statistical analysis methods.
\section*{Acknowledgments}
We would like to acknowledge G. M\'etris and L. Serron for their availability and for providing us simulated data and the related documentation.


\section*{References}

\begin{thebibliography}{99}
\bibitem{tasson2011}V. A. Kosteleck\'y and J. D. Tasson, \Journal{\PRD}{83}{016013}{2011}.

\bibitem{colladay1998}D. Colladay and V. A. Kosteleck\'y, \Journal{\PRD}{58}{116002}{1998}.

\end{thebibliography}
\end{document}